\documentclass[aps,prd,nofootinbib,twocolumn,superscriptaddress,showpacs,preprintnumbers]{revtex4}
\usepackage{amsmath,amssymb,ifpdf,url,mathrsfs,slashed,multirow,tabularx,type1cm,subfigure}
\usepackage{graphicx,color}

\ifpdf        
  \usepackage{graphics}
\else                          
 \usepackage[dvipdfmx]{graphics}      
\fi

\allowdisplaybreaks

\definecolor{BlueViolet}{rgb}{0.2, 0.00, 0.7}
\definecolor{Blue}{rgb}{0.15, 0.00, 0.9}
\usepackage[
colorlinks=true,linkcolor=Blue,citecolor=Blue, urlcolor=BlueViolet]{hyperref}

\newcommand{\GeV}{\,{\rm GeV}}
\newcommand{\MeV}{\,{\rm MeV}}

\newcommand{\Slash}[1]{{\ooalign{\hfil \hspace*{-5pt}~#1\hfil\crcr\raise.167ex\hbox{/}} }}

\def\({\left(}
\def\){\right)}
\def\<{\langle}
\def\>{\rangle}
\newcommand{\non}{\nonumber \\ }
\newcommand{\matl}{\left( \begin{array}}
\newcommand{\matr}{\end{array} \right)}

\def\beq#1\eeq{\begin{align}#1\end{align}}

\def\GeV{\text{GeV}}

\newcommand{\mcl}[1]{\mathcal{#1}}
\newcommand{\del}{\partial}
\newcommand{\n}{\notag \\}
\newcommand{\fsl}[1]{\slashed{#1}}

\newcommand{\ga}{\ensuremath{\gamma}}

\newcommand{\de}{\ensuremath{\delta}}

\newcommand{\et}{\ensuremath{\eta}}

\newcommand{\la}{\ensuremath{\lambda}}

\newcommand{\si}{\ensuremath{\sigma}}

\newcommand{\ph}{\ensuremath{\phi}}
\newcommand{\vph}{\ensuremath{\varphi}}

\newcommand{\ch}{\ensuremath{\chi}}
\newcommand{\ps}{\ensuremath{\psi}}

\begin{document}

\preprint{OU-HET-909}
\preprint{TTP16--036}

\title{Protophobic Light Vector Boson as a Mediator to the Dark Sector}

\author{Teppei Kitahara} \email{teppei.kitahara@kit.edu} 
\affiliation{
  Institute for Theoretical Particle Physics (TTP), 
  Karlsruhe Institute of Technology, Wolfgang-Gaede-Stra\ss e 1, 
  76128 Karlsruhe, Germany
}
\affiliation{
  Institute for Nuclear Physics (IKP), Karlsruhe Institute of
  Technology, Hermann-von-Helmholtz-Platz 1, 76344
  Eggenstein-Leopoldshafen, Germany
}

\author{Yasuhiro Yamamoto}
\email{yamayasu@het.phys.sci.osaka-u.ac.jp}
\affiliation{
 Department of Physics, Osaka University, Toyonaka 560-0043 Osaka, Japan
}

\date{\today}

\begin{abstract}
The observation of a protophobic 16.7 MeV vector boson has been reported by a $^8$Be nuclear transition experiment.
Such a new particle could mediate between the Standard Model and a dark sector, which includes the dark matter.
In this paper, 
we show some simple models of the dark matter which satisfy the thermal relic abundance under the current experimental bounds from the direct and the indirect detections.
In a model, it is found that an appropriate self-scattering cross section to solve the small scale structure puzzles can be achieved. 
\end{abstract}

\maketitle

\section{Introduction}
\setcounter{footnote}{0}
\renewcommand{\thefootnote}{\#\arabic{footnote}}
\setcounter{page}{1}

Although the dark matter has been gravitationally confirmed by astrophysical observations in various ways, one has no information on the properties, e.g., the mass and the coupling.
 In various dark matter models, 
a kind of popular model includes a light new boson which mediates between the Standard Model and a dark sector, e.g.,~Ref.~\cite{Pospelov:2007mp}.
Such a light particle simultaneously plays an important role in order to solve several problems, 
for instance, the small scale structure problems~\cite{Kaplinghat:2015aga}, the Lithium  problem~\cite{Goudelis:2015wpa}, and the muon $g-2$ anomaly~\cite{Endo:2012hp}. 

Recently, a $^8$Be nuclear transition experiment has reported a signal which can be interpreted as an unknown light vector boson~\cite{Krasznahorkay:2015iga}.
The vector boson ($X$) is observed as a resonance in $e^+e^-$ pairs of which the  invariant mass is $m_X = 16.7 \pm  0.35_{\textrm{stat}} \pm 0.5_{\textrm{sys}}$\,MeV.
If one supposes a vectorlike interaction between the Standard Model matter fields and the  light vector boson, the consistency in the other experimental results requires that the interaction should be protophobic~\cite{Feng:2016jff,Feng:2016ysn}, which can be written as
\begin{align}
  \mcl{L}_\text{int} =
	  -X_\mu ( g_u \bar{u} \ga^\mu u +g_d \bar{d} \ga^\mu d+g_e \bar{e} \ga^\mu e + g_\nu \bar{\nu}_L \ga^\mu  \nu_L  ),
 \label{EqInt}
\end{align}
where
\begin{align}
  & 2.0\times 10^{-4} \lesssim |g_u| \lesssim 1.0 \times 10^{-3},\n
	& 4.0\times 10^{-4} \lesssim |g_d| \lesssim 2.0 \times 10^{-3},\n
  & 6.1\times 10^{-5} <  |g_e| <  4.2\times 10^{-4},\\
  &|g_\nu g_e|^{1/2}  \lesssim 9.1\times 10^{-5}\ (\text{for } g_\nu g_e <0),\n
  &|g_\nu g_e|^{1/2}  \lesssim 2.1\times 10^{-5}\ (\text{for } g_\nu g_e >0). \notag
\end{align}
The coupling with the neutron $g_n$, which is defined as $g_n=g_u +2g_d$, satisfies $6.1\times 10^{-4}\lesssim |g_n| \lesssim 3.0\times 10^{-3}$.
On the other hand, the coupling with proton $g_p(=2g_u +g_d)$ is restricted as $|g_p| \lesssim 3.6\times 10^{-4}$.
We fix the couplings as $g_n=3.0\times 10^{-3}$, $g_p=0$, $g_e=4.2\times 10^{-4}$, and $g_\nu=0$ in the following discussion. 
The phenomenology of the $X$ boson and its  models have been investigated in Refs.~\cite{Feng:2016ysn,Gu:2016ege,Jia:2016uxs}.

\begin{figure}[tbp]
\begin{center}
 \includegraphics[width=0.35\textwidth, bb= 0 0 399 137]{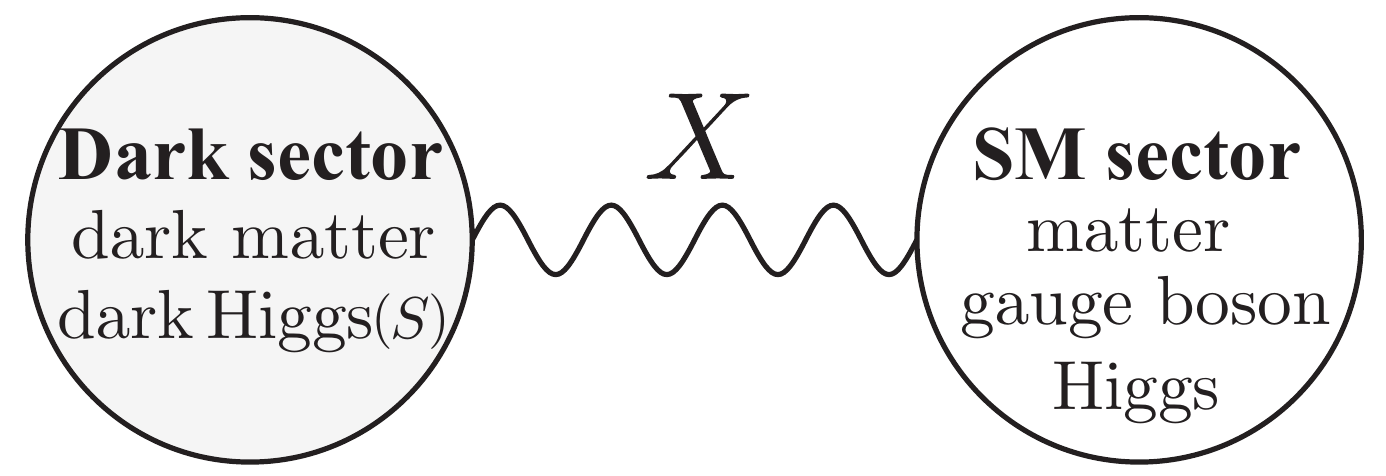}
  \caption{A schematic description of models considered in this paper.
}
\label{fig:schematic}
 \end{center}
\end{figure}

In this paper, we assume the light vector boson to be a gauge boson of a broken U(1)$_X$ gauge symmetry and to be a mediator between the Standard Model and the dark sector which includes the dark matter, as in Fig.~\ref{fig:schematic}.
Using some simple models, we investigate experimental constraints on their parameters in the dark sector, and also discuss the compatibility with the thermal relic dark matter scenario.
%

\section{Models of U(1)$_X$ charged dark matter}

For the dark sector, we consider spontaneous symmetry breaking of the U(1)$_X$ gauge symmetry by a dark Higgs $S$, which is a complex scalar boson charged under U(1)$_X$. 
The Lagrangian of the dark sector is
\begin{align}
 \mcl{L}_X =&
  -\frac{1}{4} X^{\mu\nu}X_{\mu\nu}
	+(D_\mu S)^\dag (D^\mu S) +\mu_S^2 |S|^2 -\frac{\la_S}{2} |S|^4 \non
 =&
   -\frac{1}{4} X^{\mu\nu}X_{\mu\nu} +\frac{m_X^2}{2} X^\mu X_\mu 
	 +\frac{1}{2} (\del_{\mu} s)^2 -\frac{m_s^2}{2} s^2 \n &
	 +g_X m_X s X^\mu X_\mu -\frac{g_X m_s^2}{2 m_X} s^3 +\cdots,
\end{align}
where $X^{\mu\nu}$ is the field strength tensor of $X$.
The scalar field $S$ is expanded as $S=(v_s + s)/\sqrt{2}$ in the unitarity gauge, and $D_\mu S=(\del_\mu +i g_X X_\mu) S$.
Some terms irrelevant in our computation are suppressed here.
The parameters are defined as
\begin{align}
 v_s^2 &= 2\mu_S^2 / \la_S, ~~~
 m_s^2 = \la_S v_s^2, ~~~
	 m_X = g_X v_s.
\end{align}
The original parameters $\mu_S$ and $\la_S$ can be written by $m_X$ and $m_s$ with the gauge coupling $g_X$.
For simplicity, we consider that interactions between the Higgs boson and the dark sector can be neglected.\footnote{
The couplings between the dark sector and the Higgs boson are introduced as $\la_{SH} |S|^2 |H|^2$ and $\la_{\vph H} |\vph|^2 |H|^2$.
The Higgs to the dark Higgs pair decay has four electron-positron pairs in the final state.
To suppress the significant contribution to the total width of the Higgs, since the current bound is five times larger than the Standard Model prediction, the coupling $\la_{SH}$ should be smaller than about 0.06; see Ref.~\cite{Khachatryan:2015mma,Aad:2015xua}.
The coupling $\la_{\vph H}$ is constrained by the direct detection and direct measurement of the Higgs invisible width as a Higgs portal dark matter.
Roughly speaking, the coupling should be smaller than about 0.01 to evade any experimental bound;
 see Refs.~\cite{Kanemura:2010sh,Cline:2013gha}.}

First, we study a complex scalar model  and a Dirac fermion dark matter model, where the dark matter is charged under U(1)$_X$.
If the dark matter is the complex scalar field $\vph$, the Lagrangian is
\begin{align}
 \mcl{L}_{\vph} =&
  |(\del_\mu +ig_{\vph} X_\mu) \vph|^2 -m_{\vph}^2 |\vph|^2  
 -\frac{\la_\vph}{2} |\vph|^4 \n & 
 -\la_{\vph S} |\vph|^2 |S|^2 +\mcl{L}_X.
\label{EqCdm}
\end{align}
Since the annihilations into $ss$ and $XX$ are the s-wave processes,  they dominate the thermal relic abundance.
In this model, the experimental bounds on these two channels are, in addition to the dark matter mass $m_{\vph}$, determined by the couplings $\la_{\vph S}$ and $g_{\vph}$, respectively.

\begin{figure}[tbp]
\begin{center}
 \includegraphics[width=0.4\textwidth, bb= 0 0 360 345]{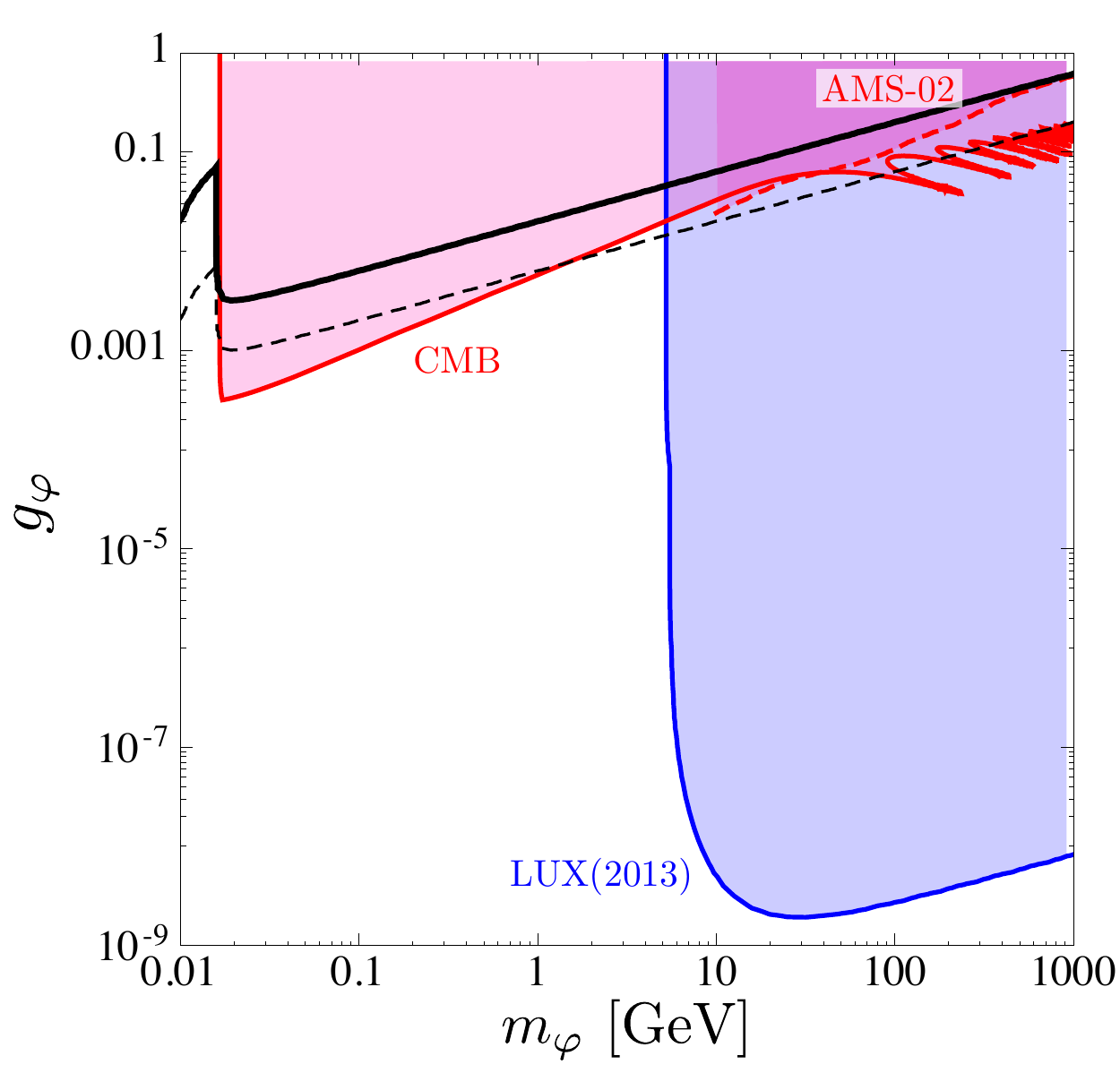}
  \caption{
  The experimental constraints on  the gauge coupling $g_{\vph}$ as a function of the dark matter $m_{\vph}$.
	The solid black line means the observed dark matter abundance, namely, $\langle \sigma v \rangle = 6 \times 10^{-26}$\,cm$^3/$s with $x=20$.
	On the dotted black line, 
	the predicted dark matter abundance is 2 orders of magnitude larger than the observed abundance.
	The red regions are excluded by the Plank and the AMS-02 experiments, while 
  the blue region has been excluded by the LUX experiment.
}
\label{fig:directcoupling}
 \end{center}
\end{figure}

The annihilation cross section at the dark age ($m_\text{DM}/T\sim 3\times 10^{12}$) is bounded by an observation of the cosmic microwave background (CMB) by the Planck as $\langle \si v \rangle /m_\text{DM} \lesssim 1.0\times 10^{-27}$cm$^3$/s/GeV~\cite{Ade:2015xua,Kawasaki:2015peu}.
Hence, the region where the dark matter is lighter than 30 GeV is naively excluded by the result.
Even if the dark matter is heavier than the value, the large Sommerfeld enhancement through the $X$ boson excludes the thermal relic scenario \cite{Hisano:2004ds, Cassel:2009wt}.
The similar bound is obtained by AMS-02 for the region $m_\vph >10\,\GeV$~\cite{Accardo:2014lma,Aguilar:2014mma} with $m_\text{DM}/T\sim 3\times 10^{6}$.
These indirect signals are one- or two-step cascades studied in Ref.~\cite{Elor:2015bho}.
The region is also excluded by the direct detection result of the LUX experiment~\cite{Akerib:2013tjd}.
To see these bounds, we have followed the analysis method used in Refs.~\cite{Gresham:2013mua,Li:2014vza,DelNobile:2015uua}.
These results are shown in Fig.~\ref{fig:directcoupling}.

For simplicity, we consider only the $XX$ channel in the figure, so that  only $g_\vph$ is a  relevant coupling.
The result can be translated into the $ss$ channel with the replacement of $g_\vph^2$ by $\la_{\vph S}/(2\sqrt{2})$ in their nonrelativistic annihilation cross sections.
Even if both of these channels contribute to the annihilation process, the thermal relic dark matter cannot be obtained.
Note that recently Ref.~\cite{Jia:2016uxs} has shown that 
if the dark matter is lighter than the vector boson a certain parameter region can explain the thermal relic abundance.

Considering the Dirac fermion dark matter $\xi$, the Lagrangian is
\begin{align}
 \mcl{L}_{\xi} =&
   \bar{\xi}(i \fsl{\del} -g_{\xi} \fsl{X} -m_{\xi}) \xi +\mcl{L}_X,
\end{align}
where $\xi$ is the dark matter.
The dark matter annihilates through the s-wave processes into $XX$ and $sX$.
The situation of the experimental bounds and the consequence for the thermal relic scenario are the same as $\vph$. 
The nonrelativistic annihilation cross section of $ \xi \bar{\xi} \to XX$ is just half of the cross section of $\vph \vph^{\ast} \to XX$   if $g_\xi = g_\vph$,
while
the result of the $sX$ channel is given by the replacement of $g_\xi^2$ with $g_\xi g_X /2$.
Since the Sommerfeld enhancement factor is the same as  $\vph$,
the result is almost the same as Fig.~\ref{fig:directcoupling}, except for the window in the light dark matter region.
Therefore, the thermal relic scenario is also excluded.

\section{Models of Secluded Dark Matter}

Next, we study the U(1)$_X$ singlet dark matter models.
Interactions between the dark matter and the Standard Model are induced by the mixing with another particle charged under the U(1)$_X$  gauge symmetry.

In the case of the real scalar dark matter $\ph$, the dark Higgs $s$ can be used as the mediator.
After the spontaneous symmetry breaking, the Lagrangian is
\begin{align}
 \mcl{L}_{\phi} =&
  \frac{1}{2} (\del^\mu \ph)^2 -\frac{m_{\phi}^2}{2} \ph^2 -\frac{\la_{\ph S} m_X }{2 g_X} s\ph^2 \n &
 -\frac{\la_{\ph S}}{4} s^2 \ph^2
 -\frac{\la_\ph}{4!} \ph^4 +  \mcl{L}_X,
\label{EqLagreal}
\end{align}
where we impose a $Z_2$ symmetry ($\phi \leftrightarrow -\phi$) to stabilize the dark matter.
The coupling $\la_{\ph S}$ is introduced like $\la_{\vph S}$.

\begin{figure}[tbp]
\centering
 \includegraphics[width=0.25\textwidth, bb= 0 0 174 106]{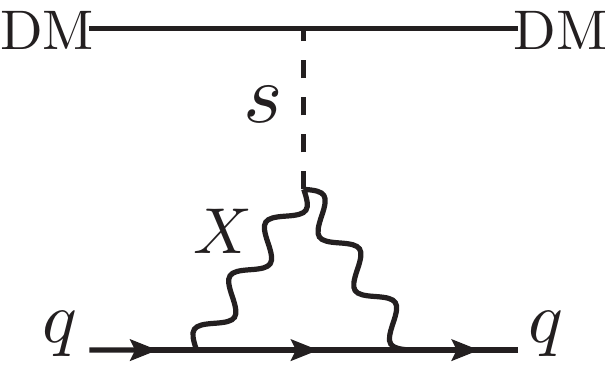}
 \caption{
 The leading contribution of the direct detection in the U(1)$_X$ singlet dark matter.
}
\label{fig:diagramdirectdetection}
\end{figure}

In the previous section, the direct detection excludes the thermal relic scenario if the dark matter is heavier than about 5\,\GeV.
 In this model, however, the leading contribution to the direct detection comes from the loop-induced diagram shown in Fig.~\ref{fig:diagramdirectdetection}.
Hence, the direct detection bound becomes significantly weaker than the previous models.

In this model, the annihilation cross section is dominated by the s-wave processes: $\phi \phi \to s^{\ast} \to XX$ and $\phi \phi \to  s s$.
Since the scalar three-point interaction is proportional to $m_X$, 
the Sommerfeld enhancement factor is also suppressed by $m_X / m_{ \phi} $ \cite{Hryczuk:2011tq}, unless $\la_{\ph S} \gg g_X$.

We also consider the U(1)$_X$ singlet dark matter model including a Majorana fermion $\ch$.
Since the Majorana fermion cannot interact with $S$ alone, we additionally introduce a Dirac fermion $\ps$ of which the U(1)$_X$ charge is the same as $S$.
Then, the Lagrangian is 
\begin{align}
 \mcl{L}_{\chi} =&
   \bar{\ps} (i\fsl{\del} -g_X \fsl{X} -m_\ps )\ps
	+\frac{1}{2} \bar{\ch}(i\fsl{\del} -m_\ch)\ch \n &
	-y (S \bar{\ps} \ch +S^\dag \bar{\ch} \ps)  +  \mcl{L}_X,
\end{align}
where the Yukawa coupling $y$ can be chosen as real and positive without loss of generality.
After the spontaneous symmetry breaking, the coupling becomes the source of the fermion mixing.
The mass eigenstates are obtained by a SO(3) rotation and a chiral rotation to flip the sign of a mass term as
\begin{align}
 \mcl{L}_{\chi} \supset &
 -\frac{1}{2} ( \bar{\ch^c}\, \bar{\ps_1^c}\, \bar{\ps_2^c} )
  \begin{pmatrix}
    m_\ch & \frac{y v_s}{\sqrt{2}} & \frac{y v_s}{\sqrt{2}} \\
    \frac{y v_s}{\sqrt{2}} & 0 & m_\ps \\ 
    \frac{y v_s}{\sqrt{2}} & m_\ps & 0
  \end{pmatrix}
	\begin{pmatrix} \ch \\ \ps_1 \\ \ps_2 \end{pmatrix}
+\text{H.c.}\non
 =&
 -\frac{1}{2} ( \bar{\et_1^c}\, \bar{\et_2^c}\, \bar{\et_3^c} )
  \begin{pmatrix}
    m_1 & 0   & 0 \\
    0   & m_2 & 0 \\ 
    0   & 0   & m_3
  \end{pmatrix}
	\begin{pmatrix} \et_1 \\ \et_2 \\ \et_3 \end{pmatrix}
+\text{H.c.}\\
= & - \frac{1}{2} m_i \, \bar{\chi}_i  \chi_i,
\end{align}
where $\ps_1$ and $\ps_2$ are, respectively, the left-handed and the charge conjugation of the right-handed components in $\ps$, namely, $\ps_2 =(\ps_R)^c$.
The mass eigenstates $\ch_i$ are the four-component Majorana fermions defined as $\chi_i = (\eta_i,\, \eta_i^c)^{T}$.
We assign the mass eigenvalues  to be $0< m_1 < m_2 < m_3$;  i.e., the dark matter is $\chi_1$.
Then, the Lagrangian is written as 
\begin{align}
 \mcl{L}_{\chi} =& 
  \frac{1}{2} \bar{\chi}_i \left( (i\fsl{\del} -m_i) \de^{ij} -g^{ij} \fsl{X} -y^{ij} s \right) \chi_j + \mcl{L}_{X},
\end{align}
where
\begin{align}
  g^{ij} =& g_X
	 \begin{pmatrix}
     0 & - i \sqrt{\frac{m_3-m_2}{m_3-m_1}} & 0 \\
     i\sqrt{\frac{m_3-m_2}{m_3-m_1}} & 0 & i\sqrt{\frac{m_2-m_1}{m_3-m_1}} \\
     0 & - i \sqrt{\frac{m_2-m_1}{m_3-m_1}} & 0
	 \end{pmatrix}, \\
  y^{ij} =& y
	 \begin{pmatrix}
    -2\frac{\sqrt{(m_3-m_2)(m_2-m_1)}}{m_3-m1} & 0 & \frac{m_3 - 2m_2 + m_1}{m_3-m_1} \\
		 0 & 0 & 0 \\
		 \frac{m_3 -2m_2 + m_1}{m_3-m_1} & 0 & 2\frac{\sqrt{(m_3-m_2)(m_2-m_1)}}{m_3-m_1}
	 \end{pmatrix}.
\end{align}
The mass eigenvalues are related as 
\begin{align}
  m_3 -m_2 = \frac{y^2 m_X^2}{g_X^2 (m_2 -m_1)}.
\end{align}
In the numerical analyses below, we chose $m_2 -m_1 = 100\,\GeV$ to evade the complexity of the coannihilation of the dark fermions.

The leading contribution to the direct detection signal is also the loop diagram given in Fig.~\ref{fig:diagramdirectdetection} like the real scalar model. 
Since the s-wave annihilation channel is suppressed by $m_X^4 / (m_2 -m_1 )^4$, the leading annihilation process is the p-wave.
Hence, the indirect detection bound does not work to exclude the thermal relic scenario for the Majorana dark matter.
The dominant contributions to the annihilation come from the annihilation into $s s$ with the $t$-channel exchange of $\chi_3$ and that into $X X$ with the $s$-channel mediation by $s$.
They are the p-wave processes.\footnotemark
\footnotetext{
Even if the annihilation is dominantly
p-wave, it has been pointed out that radiative bound state formation
can occur in the s-wave, and the resulting bound states can
subsequently decay into Standard Model particles, contributing to the
dark matter annihilation signal. This effect could change the bounds
on our models from indirect detection. However, in our scenario, the
mediator mass is much larger than the binding energy, and bound state
formation is forbidden; thus, we do not have to take this contribution
into account. The details are discussed in Ref.~\cite{An:2016gad}.
}

Even though the indirect detection bounds are too weak, the Sommerfeld enhancement factor  \cite{Hryczuk:2010zi} and the self-scattering cross section can be large if one takes the region of the heavy dark matter mass $m_1$ and the lighter mediator mass $m_s$.
In this situation, the large self-interaction can solve the small scale structure problems as shown in the next section.

Finally, we mention the compatibility of UV completions and our models.
Two models of the protophobic light vector boson have been proposed in Ref.~\cite{Feng:2016ysn}.
The vector is assigned as the gauge boson of the broken U(1)$_B$ or U(1)$_{B-L}$.
The implementation of the real scalar dark matter is straightforward.
If the scalar is heavier than the additional fermion introduced there, new annihilation channels are opened via the $s$-channel dark Higgs mediation.
In that case, the experimental constraints become weaker than our results shown in the next section.
The Majorana dark matter can be attached in the gauged U(1)$_B$ model.
However, the mixing and the masses become the same scale, namely, $m_\ch \sim m_\ps \sim v_s$.
This situation is not included in our analysis since we assume the hierarchical parameters.
Hence, the additional Dirac fermion is also required in order to apply our results to the UV completions.

\boldmath
\section{Phenomenology of the real scalar dark matter}
\unboldmath  

\begin{figure}[tbp]
\centering
 \includegraphics[width=0.4\textwidth, bb= 0 0 360 345]{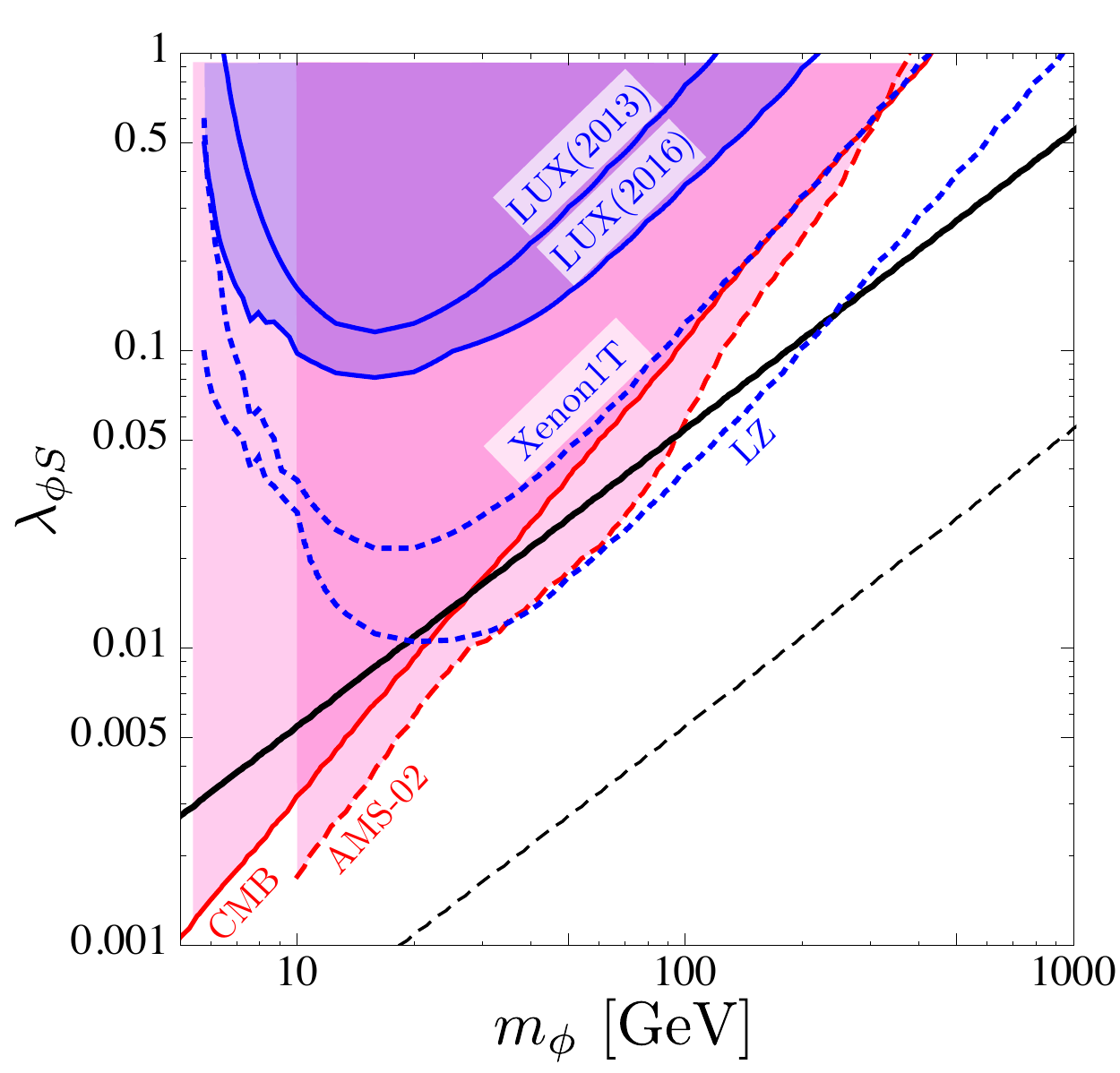}
 \caption{
  The constraints on the coupling between the dark Higgs and the dark matter $\la_{\ph S}$ as a function of the dark matter mass with $m_s=50\MeV$.
	The blue/red region is bounded by the current direct/indirect dark matter searches.
	The projected direct detection bounds by the  XENON1T and the LZ experiments are shown with the blue dotted lines.
  On the solid black line, the dark matter satisfies the observed thermal relic abundance $\langle \sigma v \rangle = 3 \times 10^{-26}$\,cm$^3/$s with $x=20$.
  On the dotted black line, 
  the predicted dark matter thermal relic density is 2 orders of magnitude larger than the observed one.
 }
\label{fig:realfig1}
\label{FigReal1}
\end{figure}
%
\begin{figure}[tbp]
\centering
 \includegraphics[width=0.4\textwidth, bb= 0 0 360 359]{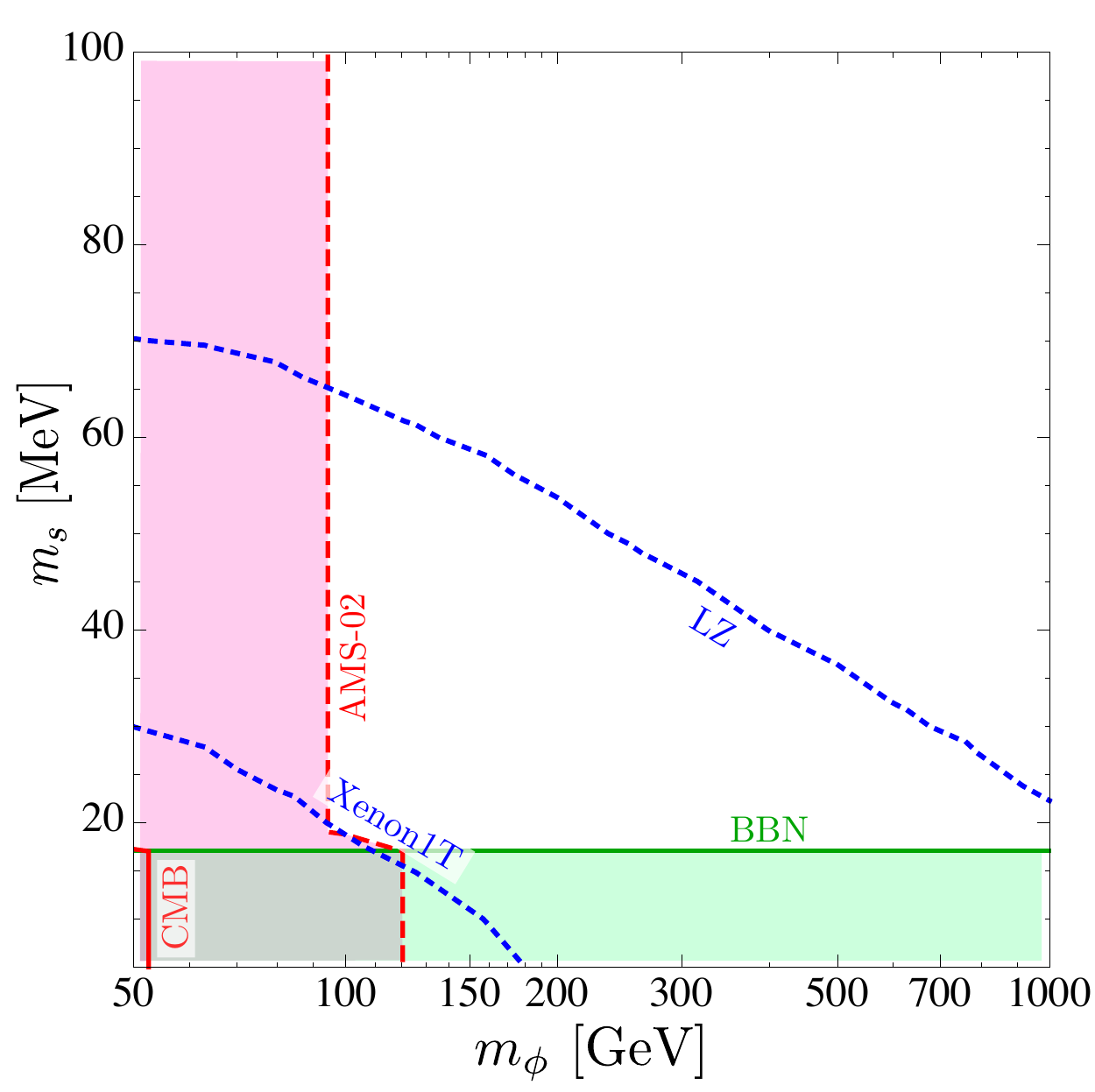}
 \caption{
  The dark Higgs mass dependence of the expected  direct detection bounds.
  The coupling $\la_{\ph S}$ is chosen to satisfy the thermal relic abundance.
  The blue dotted lines are the projected direct detection bounds.
	The red region is excluded by the indirect detections.
	In the green region, the lifetime of the dark Higgs is larger than 1 sec.
 }
\label{fig:realfig2}
\label{FigReal2}
\end{figure}

We show the indirect and the  direct detection bounds of  the real scalar dark matter model and whether they are compatible with the thermal relic abundance or not.
We also investigate future prospects of the direct detection bound.

The dark Higgs $s$ mediates the annihilation into the $X$ pair and the direct detection as Fig.~\ref{fig:diagramdirectdetection}.
In these processes, the amplitude can be written without $g_X$.
Since the annihilation to $ss$ is almost  insensitive to $g_X$, the physics of this model is described by only $\la_{\ph S}$ and the dark matter mass.
Indeed, our results are not changed in $g_X = \mathcal{O}(0.01$--$1)$.

As we have shown in Fig.~\ref{fig:realfig1},  since the Sommerfeld enhancement does not occur,  the CMB bound excludes the thermal relic scenario only if the dark matter is lighter than about 30\,\GeV.
 The bound by the AMS-02 excludes the scenario up to the mass of about 100\,\GeV.

In this paper, we consider the case in which the masses of the $X$ boson and the dark Higgs are the same scales.
Note that, because the transfer momentum in the dark matter-nucleon scattering in Fig.~\ref{fig:diagramdirectdetection} is also the same scale,
 $\mathcal{O}(10$--$100)\MeV$, 
the transfer-momentum contribution to the direct detection is not neglected. 
In addition to the  LUX bound in 2013~\cite{Akerib:2013tjd}, 
we have also drawn 
 their recent result~\cite{Akerib:2016vxi} and the prospects of the XENON1T \cite{Aprile:2015uzo} and the LZ experiments~\cite{Akerib:2015cja}.
Evaluating the hadronic matrix elements, we have used a result of the lattice QCD simulation \cite{Abdel-Rehim:2016won}.
It is found that the current LUX bound is too weak to exclude the thermal relic scenario. 
The expected bound by LZ can exclude the scenario up to a few hundred GeV, for $m_s = 50$ MeV.

The dark Higgs mass dependence of the expected direct detection bounds is shown in Fig.~\ref{FigReal2}.
It is found that 
the expected bounds by the XENON1T and the LZ experiments can exclude the scenario up to about 200 GeV and above 1 TeV, respectively.
If the dark Higgs is lighter than the $X$ boson, the scalar decays via two off-shell states.
In this case, a loop-induced decay into $e^{+} e^{-}$ becomes the dominant channel.
Eventually, the lifetime of the dark Higgs becomes larger than 1 sec.
Then,  
the observed light element abundance could be changed by large energy injection with electrons 
or inelastic scatterings between the dark Higgs and nuclei 
if the dark Higgs abundance is too much.
The direct detections can reach the higher dark matter mass for the lighter dark Higgs.
Considering the lifetime of the dark Higgs, the reaches decrease about 100 \GeV.

We also find that if one includes the strange quark contribution with $g_s = g_d$ the dark matter-nucleon scattering cross section becomes about 50\,\% larger, and the reaches increase about 100 GeV. 
The heavy quark loop contributions $(g_t = g_c= g_u,~g_b = g_d)$ are found to be  at most a 5\,\% enhancement of the scattering cross section.

\boldmath
\section{Phenomenology of the Majorana dark matter}
\unboldmath  

\begin{figure}[tbp]
\centering
  \includegraphics[width=0.4\textwidth, bb = 0 0 360 352]{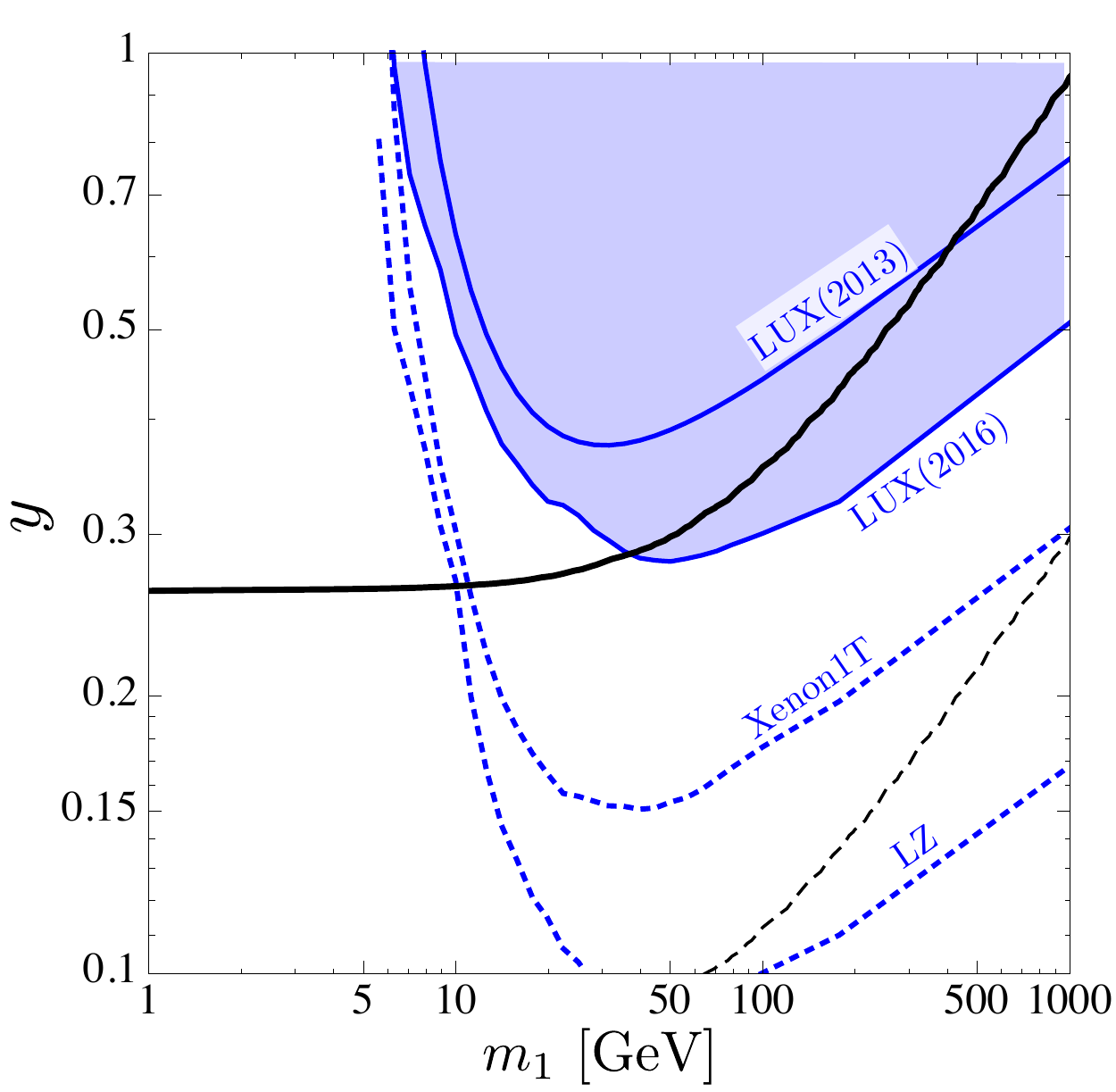}
  \caption{
	The constraints of the Yukawa coupling and the mass for the Majorana dark matter.
	The vertical axis is the Yukawa coupling $y$.
	The other objects are the same as  in Fig.~\ref{FigReal1}.
}
\label{FigMajo}
\end{figure}
%
\begin{figure}[tbh]
\centering
  \includegraphics[width=0.4\textwidth, bb = 0 0 360 356]{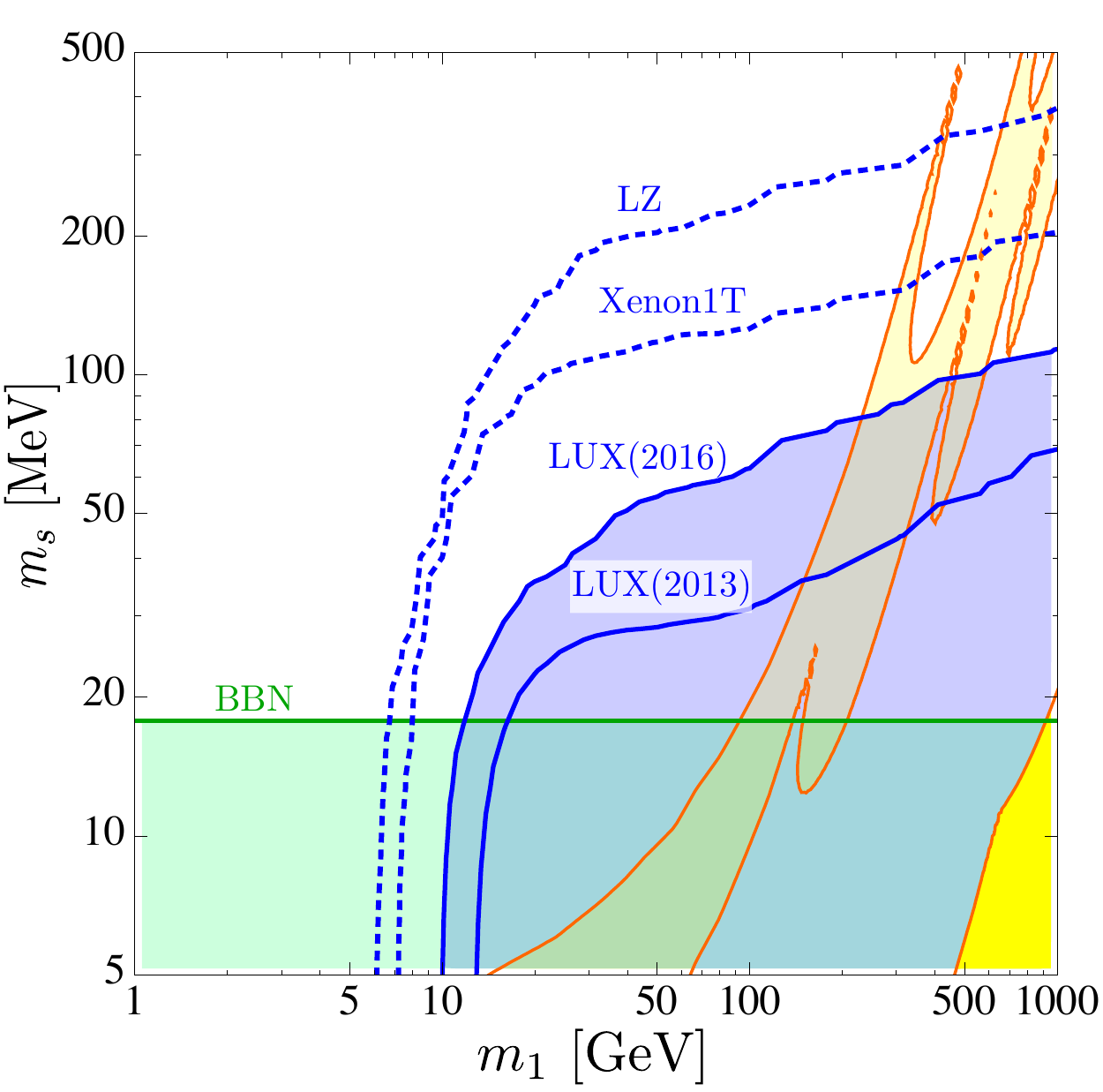}
  \caption{
	The $m_s$ dependence of the several constraints for the thermal relic Majorana dark matter.
	The self-scattering cross section to solve the small scale structure puzzle, i.e., $0.1  \leq \langle \sigma_T \rangle / m_1 \leq 10$ cm$^{2}/$g, is also shown in the yellow (lighter yellow) band for $g_X = 10^{-2}~(10^{-3})$.
	The others are the same as  in Fig.~\ref{FigReal2}.
}
\label{FigSelf}
\end{figure}

Since the Majorana dark matter mainly annihilates through the p-wave processes, only the direct detection is important to restrict the thermal relic scenario.
The bounds and the prospects are shown in Fig.~\ref{FigMajo}.

The Yukawa coupling to obtain the thermal relic abundance becomes large when the dark matter is heavier than the mass difference $m_2-m_1$.
Below the value, the cross section is determined by the mass difference, so that the coupling is independent of the dark matter mass.
The region heavier than about 40 GeV has been excluded by the direct detection for $m_s=50\MeV$.
With the projected experiments, the thermal relic region will be excluded up to the lower limit of their sensitivity, i.e., $m_1 \sim 10\,\GeV$.
Similar to the real scalar dark matter case, these behaviors are almost independent of $g_X$.

Considering the heavy dark Higgs, the direct detection bound becomes weaker.
The current bound is not sensitive if $m_s>100\,\MeV$, while
the prospected sensitivities by the  XENON1T and the LZ experiments reach the dark Higgs of 200 and 350 MeV, respectively.

The small scale structure puzzles can be solved if the velocity averaged transfer cross section of the dark matter self-scattering is as large as $\langle \si_T \rangle /m_\text{DM} \sim 0.1$--$10\text{~cm}^2/\text{g}$, see Refs.\cite{Tulin:2013teo,Kaplinghat:2015aga}.
In our Majorana model, 
the transfer cross section 
\beq
\sigma_T  = 
  \frac{3 y^8}{ \pi g_X^4} \frac{ m_X^4 m_1^2}{ m_s^4 (m_2 - m_1)^4}
\eeq
in the nonrelativistic  limit.
We find that, due to the Sommerfeld enhancement, the self-interaction can be large enough to solve the puzzles. 
In this case, the coupling $g_X$ should be smaller than $10^{-2}$.
The details are shown in Fig.~\ref{FigSelf}.

\section{Conclusion and Discussion}

In this paper, we have investigated the dark matter models where  the protophobic 16.7\,\MeV~boson is a mediator between the Standard Model and the dark sector.

Because of the severe constraint from the CMB observation  due to the large Sommerfeld enhancement, the thermal relic scenarios are almost excluded in the U(1)$_X$ charged dark matter models.

Considering  the spontaneous symmetry breaking of the U(1$)_X$ gauge symmetry,  it is found that, when the dark matter is the U(1$)_X$ singlet, the dark matter can easily satisfy the observed relic abundance under the current experimental constraints.
Some parameter regions can be probed by the future direct detection experiments.
Particularly, in the Majorana dark matter model, the large self-scattering cross section to solve the small scale structure puzzles can be achieved, while small $g_X$ is required.
These models can be easily embedded in the proposed UV completions~\cite{Feng:2016ysn}.

According to Ref.~\cite{Goudelis:2015wpa}, 
a light and a long-lived particle could solve the lithium problem.
Hence, 
the parameter region where the dark Higgs becomes a long-lived particle also attracts our attention~\cite{RefNext}.

\section*{Acknowledgments}
The work of Y.Y. is supported in part by JSPS KAKENHI Grant No. 15K05053.
The author Y.Y. thanks M. Tanaka and R. Watanabe for useful comments.

\providecommand{\href}[2]{#2}
\begingroup\raggedright

\end{document}